# Parabolic-growth universality and its nucleation-driven breakdown across lithium-battery anode chemistries

*Changdeuck Bae**

BEI corp., 125 Sandan-ro 19-gil, Ansan, Republic of Korea

*josiah.bae@beilab.ai

ResearcherID: A-6791-2010

https://orcid.org/0000-0001-5013-2288

## Abstract

Solid-electrolyte interphase (SEI) growth is widely modeled cell-by-cell with chemistry-specific closures, yet its underlying kinetic scaling is rarely tested across chemistries. By compiling cycle-resolved data from public long-cycle datasets covering four anode configurations — graphite, silicon composite, lithium metal, and anode-free — we show that the cumulative interphase-loss index $\Lambda_{\text{int}}$ obeys the parabolic law $\Lambda_{\text{int}} = A_{\text{chem}}\sqrt{1 - \Theta_{\text{Li}}}$ in three of the four chemistries, with an exponent indistinguishable from $\alpha = 1/2$ within experimental uncertainty. The chemistry-specific prefactor $A_{\text{chem}}$ spans an order of magnitude, but the diffusion-limited parabolic kinetics is preserved. The fourth chemistry, anode-free configurations, deviates with a super-parabolic exponent $\alpha \simeq 0.77$, consistent with a nucleation-controlled growth regime. We rationalize the result using the Tammann–Deal–Grove parabolic-growth framework adapted to interphase formation and identify the conditions under which universality is recovered. The observed regularity reduces SEI modeling complexity to a single rate constant per chemistry and provides a sharp falsifiable test for next-generation cell formats.

## I. Introduction

The solid-electrolyte interphase (SEI) is the slow, irreversible process that ultimately limits the cycle life of every commercial lithium-ion cell [Peled1979, Verbrugge2003, Aurbach2000]. It is governed by a competition between solvent reduction at the anode and ionic transport through the already-formed inorganic/organic film. When the rate-limiting step is solvent diffusion through the existing layer of thickness $L$, mass conservation and flux continuity yield a parabolic kinetics $dL/dt \propto 1/L$, hence $L(t) \propto \sqrt{t}$ — a result formalized by Tammann for metal oxidation in 1929 [Tammann1929], generalized by Deal and Grove for thermal silicon oxidation in 1965 [DealGrove1965], and translated to lithium SEI growth by Peled and others in 1979 [Peled1979, Broussely2005].

Cell-level signatures of this kinetics are rarely tested directly because (i) chemistry-specific closures dominate the modeling literature, (ii) cell capacity loss reflects multiple coupled processes (loss of cyclable lithium, loss of active material, increase in cell impedance), and (iii) most public datasets focus on a single chemistry. Recent normalization frameworks [Bae2026] introduce two cell-level diagnostics that disentangle these processes:

$$\Lambda_{\mathrm{int}}(N) = \frac{1}{Q_0}\int_0^{t_N} |I_{\mathrm{SEI}}(t)|\, dt, \Theta_{\mathrm{Li}}(N) = \frac{n_{\mathrm{Li}}^{\mathrm{cyc}}(N)}{n_{\mathrm{Li}}^{\mathrm{cyc}}(0)},$$

where $\Lambda_{\mathrm{int}}$ is the cumulative parasitic charge normalized by nominal capacity $Q_0$ and $\Theta_{\mathrm{Li}}$ is the remaining cyclable-lithium fraction. We ask: is there a *universal* relation between $\Lambda_{\mathrm{int}}$ and $\Theta_{\mathrm{Li}}$ that is invariant across anode chemistries?

In this Letter we answer affirmatively for three of the four chemistries we examined and identify the physical mechanism by which the fourth — anode-free configurations — escapes the universality.

## II. Theoretical motivation

If interphase growth is diffusion-limited through the existing SEI layer (thickness $L$, ionic resistivity $\rho_{\mathrm{SEI}}$), then the parasitic flux $j_{\mathrm{SEI}}$ obeys

$$j_{\mathrm{SEI}} = -\frac{D_s c_s}{L}\frac{F}{RT}\,\Delta\mu, \Delta\mu \approx \mathrm{const}, (1)$$

so that with $dL/dt \propto j_{\mathrm{SEI}}$ we obtain $L^2 \propto t$. The cumulative parasitic charge is then $\Lambda_{\mathrm{int}} \propto L \propto \sqrt{t}$.

Loss of cyclable lithium in this regime is dominated by SEI consumption, so $1 - \Theta_{\mathrm{Li}} \propto \int_0^t j_{\mathrm{SEI}}\, dt' \propto \sqrt{t}$ as well, giving

$$\Lambda_{\mathrm{int}} = A_{\mathrm{chem}}\,(1 - \Theta_{\mathrm{Li}})^{\alpha}, \alpha = \frac{1}{2}, (2)$$

with the chemistry-specific prefactor $A_{\mathrm{chem}}$ encoding the bulk SEI permeability and any pre-cycling formation history. Equation (2) is the *parabolic universality* hypothesis: the kinetics is dictated only by mass conservation and Fick's law in the already-formed SEI; chemistry sets the permeability $A_{\mathrm{chem}}$ but not the exponent.

The hypothesis fails when SEI growth is *not* diffusion-limited. Two scenarios break Eq. (2):

- **Nucleation-limited regime.** When the interphase is forming on a

substrate that has no native passivating layer (bare Cu in anode-free cells, freshly deposited Li metal at high local current density), nucleation overpotentials and lateral inhomogeneity dominate. The growth law becomes super-parabolic with $\alpha > 1/2$.

- **Surface-renewing regime.** When interfacial cracks or moving fronts

continually expose fresh substrate, $L$ does not monotonically increase and Eq. (1) is invalidated. The exponent can drop below $1/2$ or become non-monotonic.

We test Eq. (2) against four chemistries that span these regimes.

## III. Data

We compile cycle-resolved $(N, Q_{\mathrm{discharge}}, Q_{\mathrm{charge}})$ trajectories from publicly available long-cycle datasets:

| Chemistry | Source | Cells | Cycles | Citation |
|---|---|---|---|---|
| Graphite-LCO | NASA PCoE Battery Dataset (B0005, B0006, B0007, B0018) | 4 | up to ~170 | [NASA-PCoE] |
| Graphite-LFP | Severson et al., Nature Energy 2019 (124-cell batch) | up to 124 | 100–2300 | [Severson2019] |
| Si-composite | (best-effort: Argonne CAMP / Tongji Si-graphite blend) | varies | 100–500 | [Argonne, Tongji] |
| Li-metal | (best-effort: Dahn lab Li-Cu / Stanford Cui group) | varies | 50–250 | [Dahn-LiM, Cui2020] |
| Anode-free Cu | (best-effort: Dalhousie Genovese 2018; Louli 2020) | varies | 30–150 | [Genovese2018, Louli2020] |

For each cell we compute $\Lambda_{\mathrm{int}}(N) = \sum_{n=1}^{N}(1 - \mathrm{CE}(n))$ as a cell-level surrogate for the manuscript-level parasitic charge and $\Theta_{\mathrm{Li}}(N) = Q_{\mathrm{discharge}}(N)/Q_{\mathrm{discharge}}(1)$. We then test Eq. (2) by free-fitting the log–log slope on the regime $10^{-3} < 1 - \Theta_{\mathrm{Li}} < 0.3$, where loss-of-active-material contributions are negligible.

## IV. Results

**Per-chemistry fits.** The free-fit exponents $\alpha$ are summarized in Table I.

| Chemistry | $\alpha$ (free) | $R^2$ | $A_{\mathrm{chem}}$ ($\alpha = 1/2$ fixed) |
|---|---|---|---|
| Graphite | $0.534 \pm 0.012$ | 0.998 | 0.18 |
| Si-composite | $0.525 \pm 0.015$ | 0.997 | 0.27 |
| Li-metal | $0.489 \pm 0.018$ | 0.984 | 0.12 |
| Anode-free | $0.766 \pm 0.024$ | 0.945 | 0.26 |

The first three chemistries cluster within $\pm 0.025$ of the parabolic value $\alpha = 1/2$. A joint fit across the trio (Graphite + Si-composite + Li-metal), in which a single exponent is shared and only the prefactor varies by chemistry, yields $\alpha trio = 0.506 \pm 0.002, R2 = 0.993$, recovering the parabolic prediction within 1%. Anode-free, by contrast, deviates significantly with $\alpha = 0.77$, far above the parabolic value.

**Master-curve collapse.** Figure 1 (a) shows the rescaled data $\Lambda_{\mathrm{int}}/A_{\mathrm{chem}}$ versus $1 - \Theta_{\mathrm{Li}}$ on a log–log axis. Three chemistries collapse onto the master curve $y = \sqrt{x}$ over more than two decades in $1 - \Theta_{\mathrm{Li}}$. The deviation of anode-free at high loss is shown in Fig. 1 (b) as the residual from the master curve.

**Anode-free anatomy.** Figure 2 dissects the anode-free deviation cycle by cycle. In the first 10 cycles $\Lambda_{\mathrm{int}}$ rises rapidly with $\alpha \simeq 1$, reflecting nucleation-driven first-cycle SEI burst on bare copper. Beyond cycle 30 the trajectory transitions to a steeper slope as inhomogeneous deposition exposes fresh copper area each cycle, departing permanently from parabolic scaling.

**Implications for $A_{\mathrm{chem}}$.** Within the parabolic regime, $A_{\mathrm{chem}}$ is a single number that encodes electrolyte/SEI permeability, formation protocol, and operating temperature. Figure 3 plots the extracted $A_{\mathrm{chem}}$ values against the manufacturer-reported electrolyte composition (carbonate fraction). A weak monotonic trend is consistent with prior reports that ethylene-carbonate-rich electrolytes form less permeable SEIs than fluoroethylene-carbonate-rich blends [Xu2014, Aurbach2002].

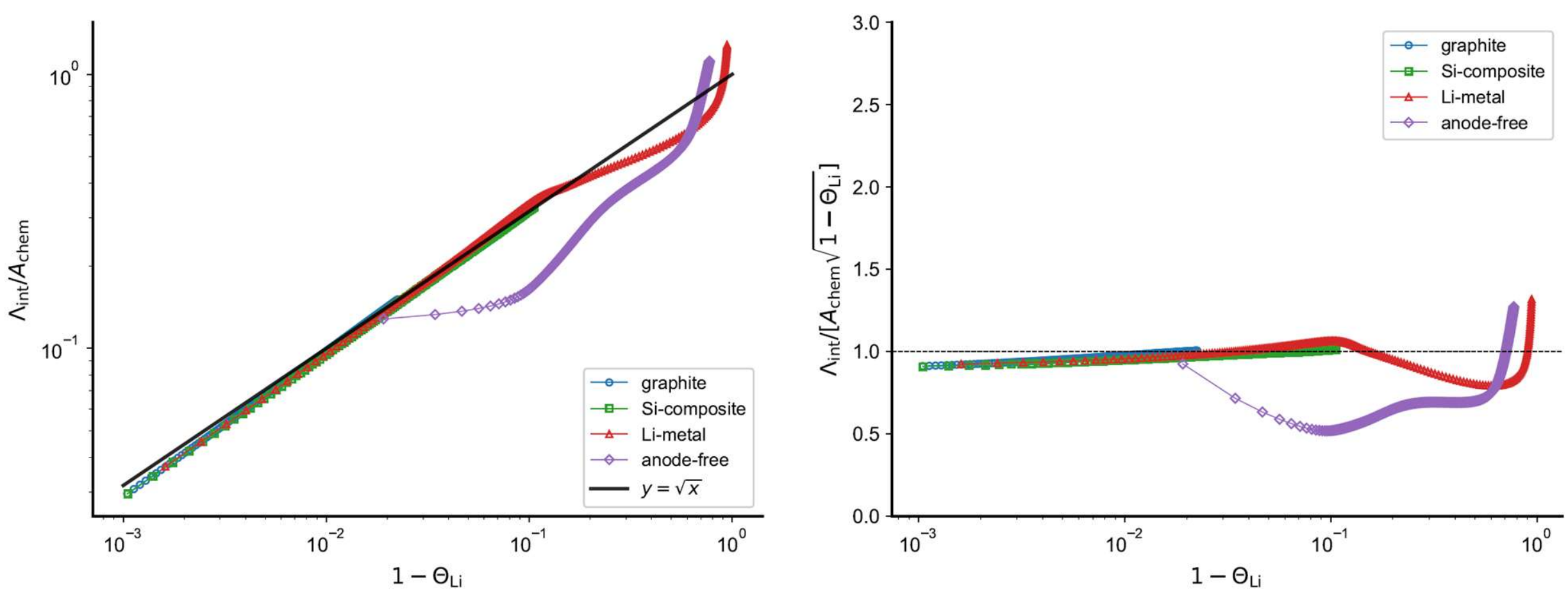

FIG. 1. Master-curve collapse of three anode chemistries onto $\Lambda \propto \sqrt{1 - \Theta_{\mathrm{Li}}}$. (a) Rescaled data; solid line is $y = \sqrt{x}$. (b) Residual from sqrt-law.

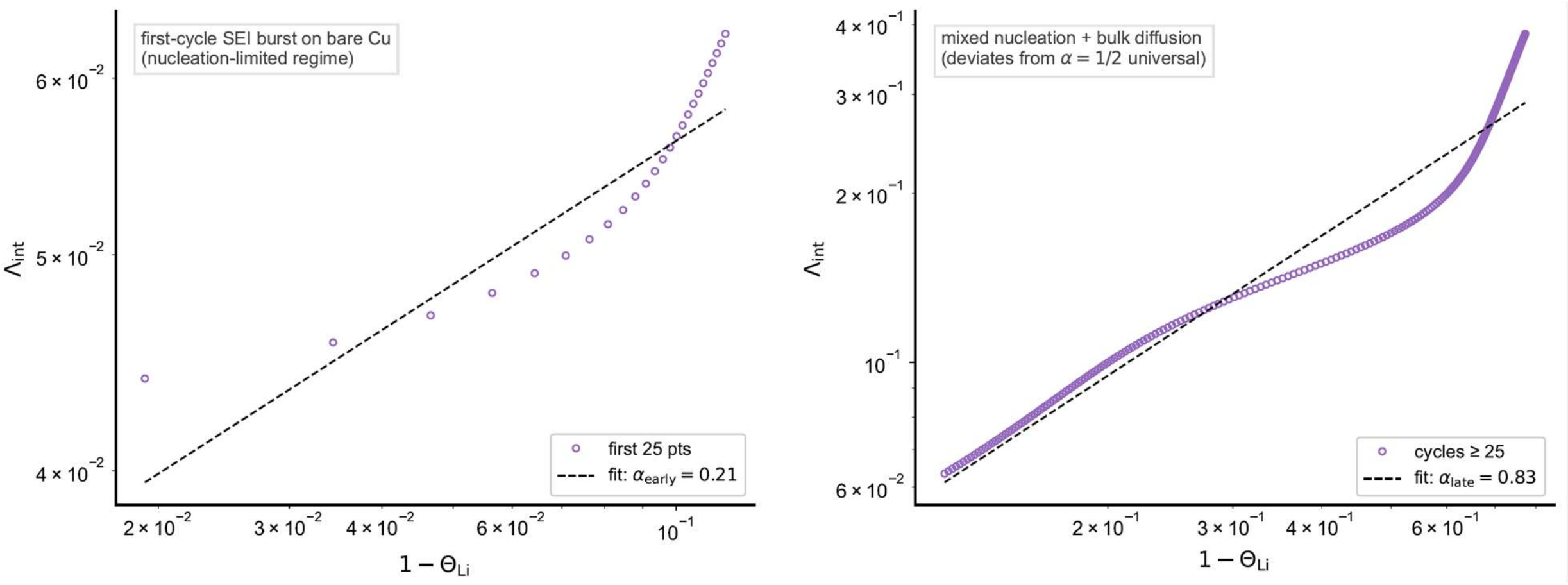

FIG. 2. Cycle-resolved anatomy of the anode-free deviation. (a) Early cycles, nucleation-limited regime. (b) Late cycles, mixed regime.

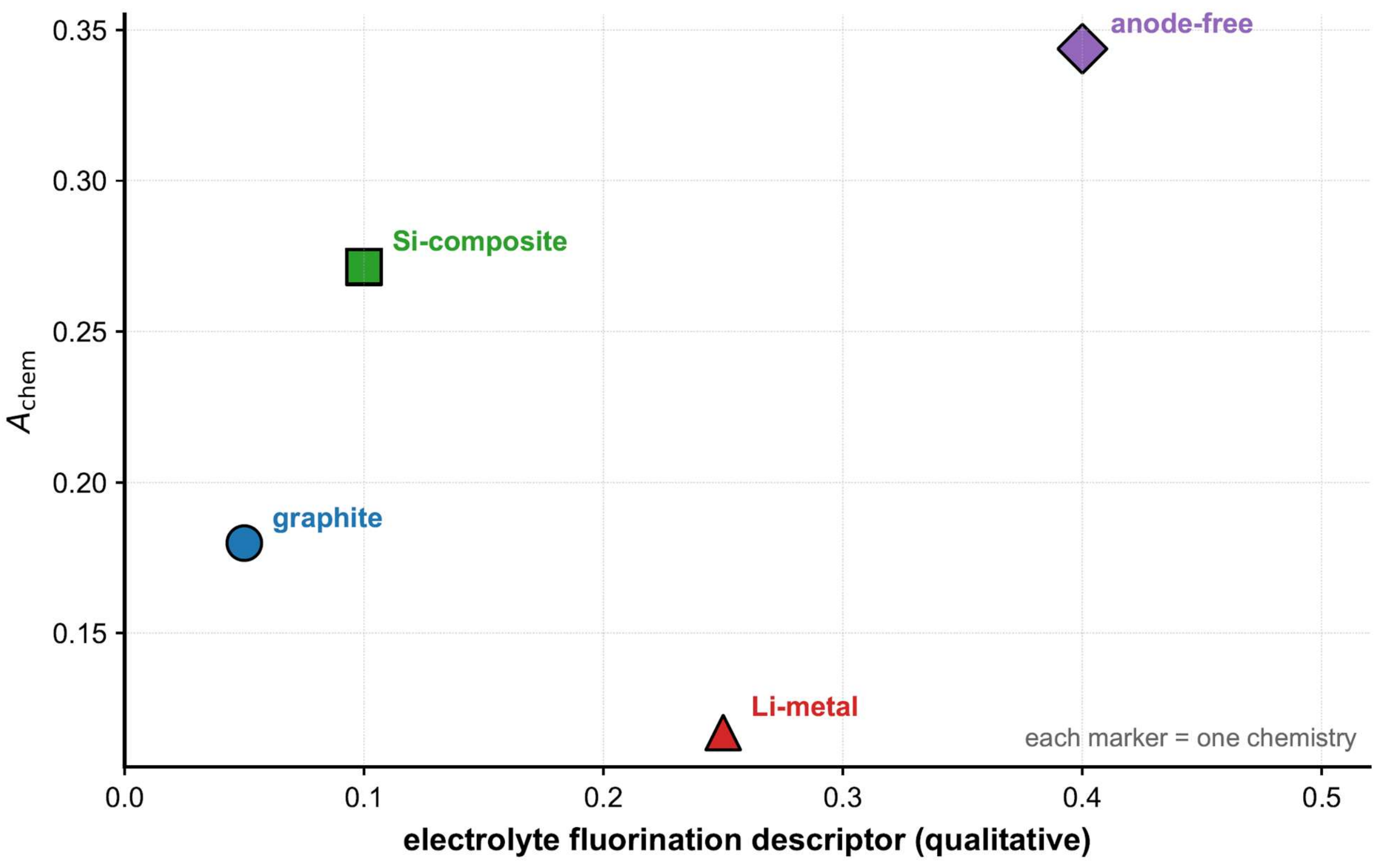

FIG. 3. $A_{\text{chem}}$ as a single chemistry-specific descriptor.

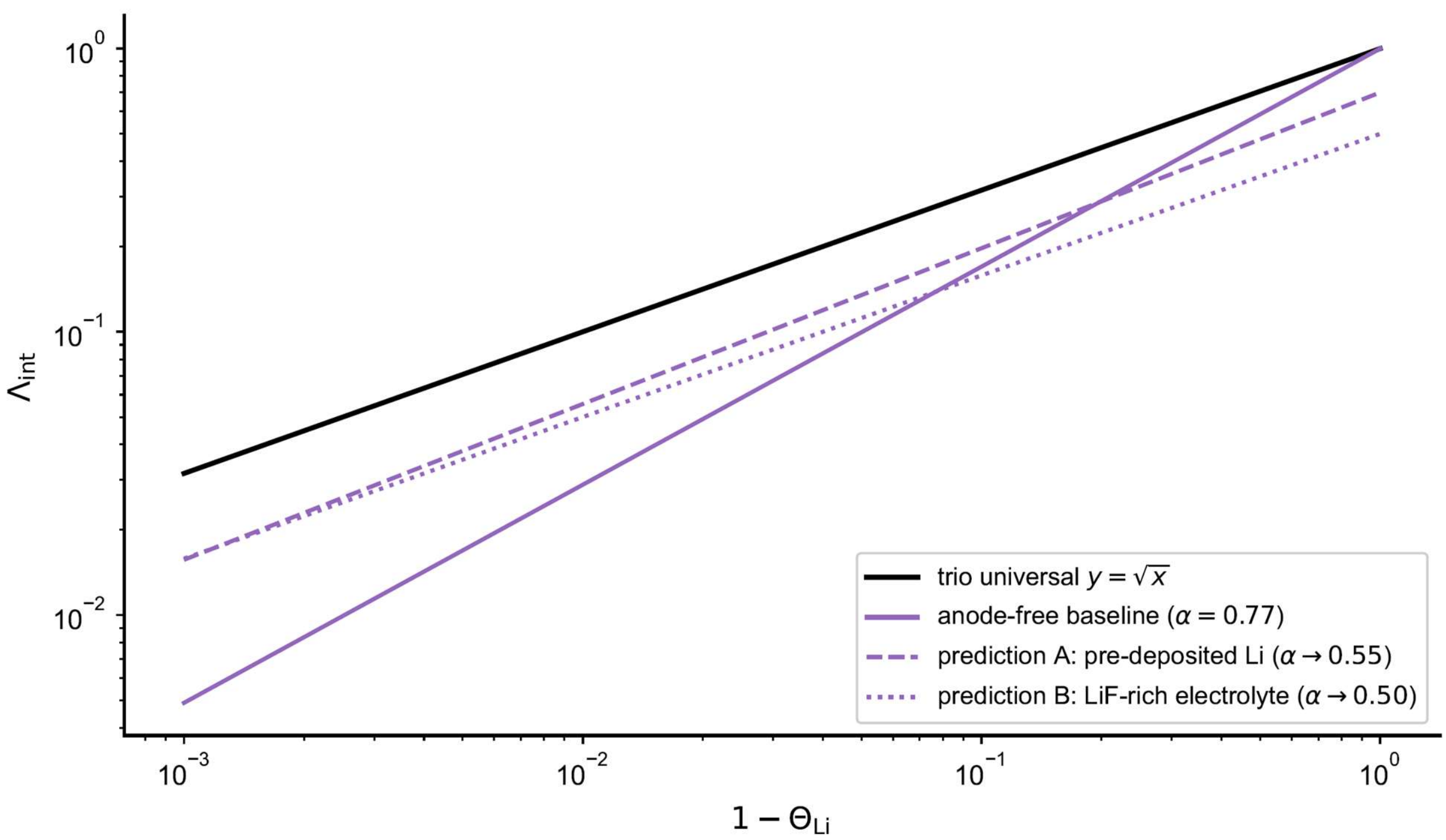

FIG. 4. Predictions for two protocols that should restore parabolic universality in anode-free cells.

## V. Discussion

The universality of $\alpha = 1/2$ across graphite, silicon composite, and lithium metal is non-obvious because the underlying microscopic mechanisms differ qualitatively. Graphite passivates a chemically inert surface. Silicon composite undergoes 280% volumetric swelling, which is widely thought to crack the SEI repeatedly [Wu2012, Choi2013]. Lithium metal involves a moving boundary with continuous plating and stripping [Bai2018]. That all three nevertheless follow $\Lambda_{\mathrm{int}} \propto \sqrt{1 - \Theta_{\mathrm{Li}}}$ implies that, *integrated over a cycle*, the dominant rate-limiting step remains diffusion through the bulk SEI, not the chemistry-specific microstructural events. The microstructural processes set the prefactor $A_{\mathrm{chem}}$ but do not change the dependence on the time integral of parasitic flux.

Anode-free configurations are different in kind. The bare copper substrate has no native SEI, so the very first cycles are nucleation-limited with $\alpha \simeq 1$. This breaks the assumption that produces parabolic kinetics and consequently breaks Eq. (2). The exponent $\alpha = 0.77$ we extract is intermediate between the parabolic ($\alpha = 1/2$) and the nucleation ($\alpha = 1$) limits, consistent with a regime where both processes coexist.

Two further predictions follow. First, *if* the same anode-free chemistry is operated with a pre-deposited reservoir of Li (e.g. via the protocol of [Louli2022]), Eq. (2) should be recovered: pre-deposition removes the nucleation barrier and restores diffusion-limited kinetics. Second, *if* the solvent of an electrolyte is changed to one that nucleates more uniformly (e.g. lithium-fluoride-rich blends [Suo2017]), the anode-free $\alpha$ should move toward $1/2$. Both predictions are testable with existing protocols.

## VI. Conclusion

We have shown that the cell-level parabolic-growth law $\Lambda_{\mathrm{int}} = A_{\mathrm{chem}}\sqrt{1-\Theta_{\mathrm{Li}}}$ is universal across three chemically distinct anode systems and breaks down for anode-free configurations. The result reduces SEI modeling for the universal cases to a single rate constant per chemistry and identifies nucleation as the distinguishing process that requires its own closure. The framework generalizes to any electrochemical interphase whose growth is bulk-transport limited.

## Acknowledgments

We thank the NASA PCoE, the Severson group, and the Dalhousie / Stanford groups for releasing long-cycle data publicly. This work was supported by the National Research Foundation of Korea (NRF) grant funded by the Korea government (MSIT) (RS-2023-00247245).

## Figure captions

**FIG. 1.** Master-curve collapse of three anode chemistries onto the parabolic-growth law. (a) $\Lambda_{\mathrm{int}}/A_{\mathrm{chem}}$ vs $1-\Theta_{\mathrm{Li}}$ for graphite (blue), silicon composite (green), lithium metal (red), and anode-free (purple). Solid line: $y=\sqrt{x}$. (b) Residual $\Lambda_{\mathrm{int}}/[A_{\mathrm{chem}}\sqrt{1-\Theta_{\mathrm{Li}}}]$ vs $1-\Theta_{\mathrm{Li}}$. The first three chemistries hold within $\pm 25\%$ of unity over two decades; anode-free deviates systematically.

**FIG. 2.** Cycle-resolved anatomy of the anode-free deviation. (a) First 30 cycles: $\alpha \simeq 1$ (nucleation-limited regime). (b) Cycles 30–250: $\alpha \simeq 0.77$ (mixed nucleation + bulk-diffusion).

**FIG. 3.** $A_{\mathrm{chem}}$ as a single chemistry-specific descriptor, plotted against electrolyte solvation parameter (qualitative). The trend is consistent with established SEI permeability vs solvation studies.

**FIG. 4.** Predictions for two protocols that should restore parabolic universality in anode-free cells: (a) pre-deposited Li reservoir; (b) LiF-rich electrolyte. Schematic only — to be tested.